\documentstyle[eqsecnum,aps]{revtex}
\begin{document}
\draft
\preprint{SOGANG-HEP 244/98, SNUTP 98-107, hep-th/9809134}
\title{Anyonic physical observables and spin phase transition}
\author{Hyun Seok Yang and Bum-Hoon Lee}
\address{Department of Physics, Sogang University, Seoul 121-742, Korea}
\maketitle

\begin{abstract}
The quantization of charged matter system coupled to Chern-Simons 
gauge fields is analyzed in a covariant gauge fixing, and gauge invariant 
physical anyon operators satisfying fractional statistics are constructed 
in a symmetric phase, based on Dirac's recipe performed on QED. 
This method provides us a definite way of identifying physical spectrums 
free from gauge ambiguity and constructing physical anyon operators 
under a covariant gauge fixing. We then analyze the statistical spin phase 
transition in a symmetry-broken phase and show that
the Higgs mechanism transmutes an anyon satisfying 
fractional statistics into a canonical boson, 
a spin 0 Higgs boson or a topologically massive photon.  
\end{abstract}

\section{INTRODUCTION}

In (2+1)-dimensional field theories such as Chern-Simons gauge theory 
with a charged matter field \cite{Hage84}, nonlinear $\sigma$ 
model with a Hopf term \cite{WZ83} and $CP^1$ model 
with a Chern-Simons term \cite{Poly88}, it has been known that there exist 
excitations with exotic statistics - anyons \cite{Seme88,Forte}, 
which continuously interpolate between bosons and fermions. 
The fundamental role of fractional statistics in 
condensed matter physics has been proposed 
in order to describe a fractional quantum Hall effect \cite{Laug83} 
and the behavior of two-dimensional materials such as vortices in 
superfluid helium films \cite{Hald85} and the $Cu$-$O$ planes of 
the cupper-oxide superconductors \cite{Science}. 

However, in (2+1)-dimensional quantum field theory, the explicit construction 
of anyon operators exhibiting fractional statistics has led much 
controversy and debate \cite{Hage89}. 
Although the physical results must be independent of a gauge 
fixing condition, different results with different gauge choices had 
been reported due to the difficulty of identifying 
physical degrees of freedom \cite{Foer89}. 
The gauge independent analysis was also 
attempted \cite{Bane92,Boya92,park2}.

Since the representation of a {\it physical} operator \cite{Dirac58} can be 
varied each gauge fixing, the construction of a physical operator 
under a gauge fixing condition must be treated carefully. 
One of the reasons for consideration of {\it physical} 
field variables is that the formulation of dynamics in terms of 
a Lagrangian (or Hamiltonian) and the equations of motion make use 
of a larger field algebra which includes non-observable 
fields \cite{Strocchi}. 
For example, the quantization under a covariant gauge fixing
\cite{mrfj,shin-brst} requires a consistent truncation into a physical
Hilbert space or Fock space only constructed by physical operators. 
It is important to notice that gauge invariance of an operator 
does not necessarily imply it to be {\it physical}. 
An additional care must be 
paid to identify the physical spectrums 
with correct quantum numbers such as spin and charge. 

There are also different opinions on the anyonic properties including 
the existence of statistical spin phase transition in a 
symmetry-broken phase \cite{Wen89,Boya90,shin-ssb}, which is an interesting 
problem that may be relevant to high-$T_c$ 
superconductivity \cite{WZ89}. 
There are arguments that 
the excitations relative to ground state 
are spin 0 bosons \cite{Boya90} and 
that the statistics-changing phase transition is impossible in
the Chern-Simons-Higgs theory \cite{shin-ssb}. 
These results seem to contradict the result of Deser and 
Yang \cite{Dese89}, observing that Higgs mechanism transmutes a 
non-dynamical Chern-Simons term into topologically massive, 
parity-violating, spin 1 theory \cite{Dese84}. 

In this paper, we will perform a careful analysis on 
the charged matter fields coupled to Chern-Simons 
gauge fields in a covariant gauge along the Dirac's method 
performed on QED \cite{Dirac58}. 
In Sec.II, we will analyze the Maxwell theory both in a symmetric and 
in a symmetry-broken phase with the covariant gauge fixing 
in order to illustrate the definite way of identifying physical spectrum 
free from gauge ambiguity arising from gauge fixing 
and to manifest the importance of consideration of 
{\it physical} fields. As a result, we show that there exists a spin
phase transition in the Maxwell-Higgs theory. 
In Sec.III, the quantization of Chern-Simons matter system will be 
presented and physical anyon operators will be constructed in a 
covariant gauge based on the approach in Sec.II. 
We will also analyze the quantization of the Chern-Simons matter system 
in the symmetry-broken phase and show that Higgs mechanism 
transmutes an anyon satisfying 
fractional statistics into a spin 0 Higgs boson and a topologically massive 
photon which is a Chern-Simons gauge field absorbed a would-be-Goldstone 
boson. Thus the Higgs effect influences the spin phase of the anyon 
and interestingly induces the statistical spin phase transition, 
as predicted by Wen and Zee \cite{Wen89}. 
This result is consistent with the observation of 
Deser and Yang \cite{Dese89}.   
Section IV contains conclusion. 
In Appendix, we will analyze the Poincar\'e algebra of massive vector 
fields - Proca and massive Chern-Simons theories - and 
extract the spin content 
of massive vector fields with no ambiguity.

\section{MAXWELL THEORY IN A COVARIANT GAUGE}
\subsection{Symmetric Phase}

\def\be{\begin{equation}}
\def\ee{\end{equation}}
\def\bea{\begin{eqnarray}}
\def\eea{\end{eqnarray}}
\def\ba{\begin{array}}
\def\ea{\end{array}}
\def\l{\label}
\def\c{\c}
In this subsection, we will briefly review the analysis of the Maxwell 
theory subject to a covariant gauge fixing in a symmetric phase.
The Lagrangian is  given by 
\bea
\label{lag1}
\ba{l}
{\cal L}= -\frac{1}{4}F_{\mu\nu}F^{\mu\nu}+ |D_{\mu}\phi|^2
+{\cal L_{GF}}, \\ 
{\cal L_{GF}}=-\frac{1}{2}(\partial_{\mu}A^{\mu})^2, 
\ea
\eea
where $D_{\mu}= \partial_{\mu}+ieA_{\mu}$ is a covariant derivative with 
metric $g_{\mu\nu}=(+1,-1,-1)$. 
The equal-time commutators are as follows 
(a time argument of operators is suppressed):
\bea
\label{comm1-3}
\ba{lll} 
&[A_{\mu}({\bf x}), A_{\nu}({\bf y})]=
[E_{\mu}({\bf x}), E_{\nu}({\bf y})]=0, \\
& [E_{\mu}({\bf x}), A_{\nu}({\bf y})]=ig_{\mu\nu}\delta^{2}
({\bf x} - {\bf y}), \\ 
& [\phi({\bf x}), \pi({\bf y})]=
[\phi^{\ast}({\bf x}), \pi^{\ast}({\bf y})]=
i\delta^{2}({\bf x} - {\bf y}), 
\ea
\eea
where we put 
\[E_{\mu}({\bf x})=\frac{\partial A_{\mu}}{\partial x_{0}},\;\;
\pi({\bf x})=(D_{0}\phi)^{\ast}({\bf x}),\;\;  
\pi^{\ast}({\bf x})=(D_{0}\phi)({\bf x}). \]

In order to get consistent quantum theory in the covariant gauge, 
we must impose a supplementary condition for any state $|P>$ in an 
indefinite state space \cite{Dirac58},   
\be
\label{F}
\partial_{\mu}A^{\mu}(x)|P>=0\;\;\Rightarrow 
\partial_{\mu}A^{\mu} \approx 0. 
\ee
We also require that this condition is satisfied afterwards:
\be
\label{G}
\partial_0\partial_{\mu}A^{\mu} \approx 0.
\ee
According to the arguments of Dirac \cite{Dirac58}, one can easily see the 
equations (\ref{F}) and (\ref{G}) are the only independent supplementary 
conditions affecting dynamical vaiables at one instant of time.
Then the state $|P>$ satisfying the conditions (\ref{F}) and (\ref{G}) 
is defined as {\it physical} state. 
The condition for a dynamical variable $\Phi$ to be {\it physical} is that
\be
\label{physc}
[F, \Phi] \approx 0
\ee
for each supplementary condition $F|P>=0$.

To find the physical fields, let us 
decompose $A_i(\bf x)$ and $E_i(\bf x)$ into transverse and 
longitudinal parts, 
\be
\label{tlda}
A_i({\bf x})={\cal A}_i({\bf x})+\frac{\partial V}{\partial x^i}
\ee
with $\frac{\partial {\cal A}^i}{\partial x^i}=0$ and
\be
\label{tlde}
E_i({\bf x})={\cal E}_i({\bf x})+\frac{\partial U}{\partial x^i}
\ee
with $\frac{\partial {\cal E}^i}{\partial x^i}=0$ and 
$ U=\frac{\partial V}{\partial x^0}$.
Observe that, for massless theory, the transverse-longitudinal 
decompositions, Eqs. (\ref{tlda}) and (\ref{tlde}), are free from 
the ambiguity existing in zero-momemtum limit 
for massive theory \cite{Dese82} and the Poincar\'e algebra 
in terms of the physical variables is well defined. 
Indeed, this situation corresponds to the vanishing of topological 
Chern-Simons term, $\mu=0$, in Ref. \cite{Dese82}, which shows that 
its Poincar\'e algebra is naturally free from zero-momentum anomaly. 
The conditions (\ref{F}) and (\ref{G}) can then be rewritten as
\bea
\label{F1}
&&F\equiv\partial_{\mu}A^{\mu}=E_0-\nabla^2 V\approx 0, \\
\label{G1}
&&G\equiv\partial_i F^{io}-J_0 =\nabla^2(U-A_0)-J_0\approx 0, 
\eea
where charge density $J_0(\bf x)$ is given by
\[ J_0({\bf x})=ie\left\{\pi^{\ast}({\bf x})\phi^{\ast}({\bf x})- 
\pi({\bf x})\phi({\bf x})\right\}. \] and $G$ is a Gauss-law constraint. 

From the commutation relation (\ref{comm1-3}), one can obtain a useful 
relation \cite{Dirac58} 
\be
\label{uv2}
[U({\bf x}), V({\bf y})]=iG({\bf x} - {\bf y}),
\ee
where $G({\bf x} - {\bf y})$ is a two-dimensional Green's function
\[\nabla^2G({\bf x} -{\bf y})=\delta^{2}({\bf x} - {\bf y}),\;\;\;\;\;
G({\bf x} - {\bf y})=\frac{1}{2\pi}ln|{\bf x} - {\bf y}|.\] 

Then one can see that the transverse field variables ${\cal A}^i$ and 
${\cal E}^i$ evidently commute with the supplementary conditions 
(\ref{F1}) and (\ref{G1}) and so are physical, 
while the operators $\phi$ and $\phi^{\ast}$ are unphysical. 
Define 
\be
\label{dphi1}
\tilde\phi(x)=e^{ieV(x)}\phi(x).
\ee
$\tilde\phi(x)$ now commutes with $F$ and $G$ and hence is physical. 
Similarly, $\tilde\phi^{\ast}(x)$ is physical. 
In the covariant gauge, 
the physical operators $\tilde\phi$ and $\tilde\phi^{\ast}$ describe 
charged particles surrounded by their Coulomb fields. 
Accordingly, the composite nonlocal operators
$\tilde\phi$ and $\tilde\phi^{\ast}$ just correspond to the physical processes 
of creation and annihilation of charged particles, since these processes 
must always be accompanied by an appropriate Coulomb change in an 
electric field around the point where the particle is created or 
annihilated \cite{Dirac58}. The non-locality of the physical field 
variables $\tilde\phi$ and $\tilde\phi^{\ast}$ are not surprising since 
a field carrying a non-zero charge whose current obeys 
a local field equation, $\partial_{\nu} F^{\nu\mu}=J^{\mu}$, 
cannot be local \cite{Strocchi}. 

The covariant derivatives of
$\phi$ and $\phi^{\ast}$ can then be represented by the physical fields
$\tilde\phi$ and $\tilde\phi^{\ast}$:
\bea
\label{tphy}
\ba{llll}
& D_{i}\phi(x) = e^{-ieV(x)}(\partial_{i}+ie{\cal A}_{i})\tilde\phi(x)
\equiv e^{-ieV(x)}\tilde D_{i}\tilde\phi(x),\\
& \pi(x) = e^{ieV(x)}\left\{\partial_{0}-ie(A_{0}-U)\right\}
\tilde\phi^{\ast}(x)\equiv e^{ieV(x)}\tilde\pi(x),\\
& (D_{i}\phi(x))^{\ast} = e^{ieV(x)}
(\partial_{i}-ie{\cal A}_{i})\tilde\phi^{\ast}(x)\equiv 
e^{ieV(x)}(\tilde D_{i}\tilde\phi(x))^{\ast},\\
& \pi^{\ast}(x) = e^{-ieV(x)}\left\{\partial_{0}+ie(A_{0}-U)
\right\}\tilde\phi(x) \equiv e^{-ieV(x)}\tilde \pi^{\ast}(x).
\ea
\eea
The variables ${\cal A}^i$, ${\cal E}^i$, $\tilde\phi$, 
$\tilde \pi$, $\tilde\phi^{\ast}$, and $\tilde\pi^{\ast}$ 
are the only independent physical variables, apart from the 
quantities (\ref{F1}) and (\ref{G1}). 
Now we can find the relationships between Coulomb gauge and covariant gauge 
in the state space constructed by only physical variables. 
The  field variables in covariant gauge ${\cal A}_i = A_i - \partial_i V$, 
$A_0 - U$, $\tilde\phi$, and $\tilde\phi^{\ast}$ 
correspond to $A_i$, $A_0$, $\phi$, and $\phi^{\ast}$ 
in Coulomb gauge, respectively. 
Indeed, we can reexpress the commutation 
relations (\ref{comm1-3}) as those of the physical 
field variables using the relation (\ref{uv2}):    
\bea
\label{ccomm}
\ba{llll}
& [{\cal A}_i ({\bf x}),{\cal A}_j ({\bf y})]=
 [{\cal E}_i ({\bf x}),{\cal E}_j ({\bf y})] = 0, \\
& [{\cal E}_i ({\bf x}),{\cal A}_j ({\bf x^{\prime}})]=-i
\delta^{tr}_{ij}({\bf x} - {\bf y}),\\  
& [\tilde\phi({\bf x}),\tilde\pi({\bf y})] =
[\tilde\phi^{\ast}({\bf x}),\tilde\pi^{\ast}({\bf y})] =
i\delta^{2}({\bf x} - {\bf y}), \\
& [\tilde\phi({\bf x}),{\cal A}_{i}({\bf y})] =
[\tilde\phi({\bf x}),{\cal E}_{i}({\bf y})] = 0,
\ea
\eea
where $\delta^{tr}_{ij}({\bf x} - {\bf y})$ is given by
\[ \delta^{tr}_{ij}({\bf x} - {\bf y})=
\int \frac{d^2 k}{(2\pi)^2}e^{i{\bf k}\cdot({\bf x} - {\bf y})}
(\delta_{ij}-\frac{k_i k_j}{{\bf k}^2}). \]
Also using the relation (\ref{uv2}), we obtain
\[[A_0 ({\bf x})-U({\bf x}), \tilde\phi({\bf y})]=
eG({\bf x} - {\bf y})\tilde\phi({\bf y}).\]
These are exactly the same as the commutation relations of Coulomb gauge 
in the presence of interaction \cite{Bjor65}. 

For simplicity, we can drop the gauge fixing term in the 
Langrangian (\ref{lag1}) without loss of generality 
since we impose the supplementary conditions (\ref{F1}) and (\ref{G1}) 
for physical states and all the dynamical and physical variables commute 
with these conditions. 
If we keep the gauge fixing term in the Lagrangian (\ref{lag1}), 
we will only obtain the {\it weakly} identical results after working 
out all the commutator algebra. 

Let us define an angular 
momentum operator constructed from the gauge invariant, symmetric 
energy-momentum tensor defined as
\be
\label{emt}
\delta S=\frac{1}{2} \int_{\cal M} \sqrt g 
\delta g^{\mu\nu} T_{\mu\nu}. 
\ee 
The energy-momentum tensor $T_{\mu\nu}$ for the Langrangian (\ref{lag1}) 
is then given by
\be
\label{T1}
T_{\mu\nu} =-F_{\mu\lambda}{F_{\nu}}^{\lambda} 
+(D_{\mu}\phi)^*D_{\nu}\phi+(D_{\nu}\phi)^*D_{\mu}\phi
-g_{\mu\nu} \left\{-\frac{1}{4}F_{\lambda\rho}F^{\lambda\rho}+
|D_{\lambda}\phi|^2 \right\}.
\ee
Our interest is the rotational property of physical fields. 
In order to calculate the angular momentum operator $L$, we begin 
with the symmetric energy-momentum tensor in Eq. (\ref{T1}):
\bea 
\label{L1}
L&=&\int d^2 y \epsilon^{ij} y_i T_{0j}\nonumber \\
&=&\int d^2 y \epsilon^{ij} y_i \{\tilde\pi({\bf y})
\partial_{j}\tilde\phi({\bf y}) + \tilde\pi^{\ast}({\bf y})
\partial_{j}\tilde\phi^{\ast}({\bf y})\}
-\int d^2 y \epsilon^{ij} y_i {\cal A}_{j}({\bf y})J_0({\bf y})\nonumber\\
&+&\int d^2 y \epsilon^{ij} y_i \{E^k({\bf y})-\partial^k A_0({\bf y})\}
\{\partial_k {\cal A}_{j}({\bf y})-\partial_j
{\cal A}_{k}({\bf y})\}\nonumber\\
&=&\int d^2 y \epsilon^{ij} y_i \{\tilde\pi({\bf y})
\partial_{j}\tilde\phi({\bf y}) + \tilde\pi^{\ast}({\bf y})
\partial_{j}\tilde\phi^{\ast}({\bf y})\}
+\int d^2 y \epsilon^{ij} y_i \dot{\cal A}_k({\bf y}) \partial_j
{\cal A}_{k}({\bf y})-\int d^2 y \epsilon^{ij}
\dot{\cal A}_i({\bf y}) {\cal A}_{j}({\bf y}).
\eea
In order to the final expression (\ref{L1}), we used the Gauss-law 
constraint (\ref{G1}). Due to a particular feature in three dimensions, 
the last term in Eq. (\ref{L1}), would-be-spin term, identically vanish 
since the massless vector theory has only one degree of freedom so 
that there is no additional degree of freedom available to form non-zero 
spin states (if we take ${\cal A}_i=\epsilon^{ij} \partial_j \xi$, 
$\epsilon^{ij}\dot{\cal A}_i{\cal A}_j$ is then total derivative). 
Although the second term in Eq. (\ref{L1}) does not verify the 
commutation relation characteristic of angular momentum 
(remember $[{\cal A}_i ({\bf x}),\dot{\cal A}_j ({\bf y})]=i
\delta^{tr}_{ij}({\bf x} - {\bf y})$), it can be shown that 
this term is independent of the possible polarizations of photons, 
so purely ``orbital'' in terms of an appropriate choice of 
the polarization vectors satisfying transversality condition. 
Thus a massless vector theory in three dimensions is spinless, 
confirming the result of Binegar \cite{Bine}. 

Note that one can make use of 
supplementary conditions only after we have worked out 
all the commutator \cite{Dirac64}. Following this rule, we get
\bea
\label{cang1}
[L, \tilde\phi({\bf x})]&=& 
\int d^2 y \epsilon^{ij} y_i [\tilde\pi({\bf y})
\partial_{j}\tilde\phi({\bf y}) + \tilde\pi^{\ast}({\bf y})
\partial_{j}\tilde\phi^{\ast}({\bf y}), \tilde\phi({\bf x})]\nonumber \\
&-&\int d^2 y \epsilon^{ij} y_i {\cal A}_{j}({\bf y})[J_0({\bf y}), 
\tilde\phi({\bf x})] +\int d^2 y \epsilon^{ij} y_i 
[E^k({\bf y}), \tilde\phi({\bf x})] 
\left\{\partial_k {\cal A}_{j}({\bf y})-\partial_j 
{\cal A}_{k}({\bf y})\right\} \nonumber \\
&=&-i\epsilon^{ij} x_i \partial_j\tilde\phi({\bf x}).
\eea
Since the angular momentum operator is Hermitian, 
we also obtain
\be
\label{cang2}
[L, \tilde\phi^{\ast}({\bf x})]=
-i\epsilon^{ij} x_i \partial_j\tilde\phi^{\ast}({\bf x}).
\ee
These are natural results since there is no 
anomalous spin in the Maxwell theory. 

However, if we do not use the physical fields commuting with supplementary 
conditions, we will have extra terms which depend on gauge fields. 
They will make the physical interpretation 
about rotational property of fields obscure. 
This suggests that one should first solve 
constraints (\ref{F1}) and (\ref{G1}), i.e. one has to find all 
physical objects before turning to dynamics and provides the
importance of the physical variables.        

\subsection{Symmetry-Broken Phase}

For the Maxwell theory of symmetric phase, our prescription of 
physical variables on gauge fields are based on 
transverse-longitudinal decomposition of the fields. 
However, in the case of symmetry-broken phase, we will be faced with 
a problem, zero-momentum ambiguity in 
the transverse-longitudinal decomposition of gauge fields. 
This ambiguity interrupts us from defining the Poincar\'e algebra 
in terms of physical variables and thus extracting the spin content 
of gauge fields \cite{Dese82}. 
Thus we will use two alternative and complementary prescriptions 
in order to identify physical spectrums.
  
To begin with, we introduce a 
symmetry breaking potential $V(\phi)$ in the Lagrangian (\ref{lag1}) 
and, for definite physical spectrums, consider the following 
parameterization of a charged scalar field $\phi(x)$:
\be
\label{dphi2}
\phi(x)=\frac{1}{\sqrt 2}\{v+\varphi(x)\}e^{i\chi(x)/v},
\ee
where vacuum expectation value $v$ of $\phi$ is non-zero. 
Note that, in the symmetric phase with $v=0$, the theory using 
this reparameterization becomes singular at $\varphi(x)=0$ 
since the Hessian determinant vanishes there \cite{Gitm90} 
and thus one has to be very careful in the analysis.
The Lagrangian (\ref{lag1}) with the potential $V(\phi)$
now becomes in terms of the Higgs field $\varphi(x)$ 
and the would-be-Goldstone boson $\chi(x)$
\be
\label{lag2}
{\cal L}= -\frac{1}{4}F_{\mu\nu}F^{\mu\nu}+
\frac{1}{2}(\partial_{\mu}\varphi)^2+\frac{1}{2}e^2 
(v+\varphi)^2\bar{A}_{\mu}^2-V(v,\varphi)+ {\cal L_{GF}}, 
\ee
where $\bar{A}_{\mu}$ is defined by
\be
\label{bara1}
\bar{A}_{\mu} \equiv A_{\mu}+\frac{1}{m}\partial_{\mu}\chi,\;\;\;\;\;m=ev.
\ee
The Lagrangian (\ref{lag2}) has no constraint - nonsingular theory, 
so the canonical quantization is straightforward.
From the Lagrangian (\ref{lag2}), one can obtain the conjugate momenta 
$\pi_{\varphi}$ and $\pi_{\chi}$ of $\varphi$ and $\chi$
\[ \pi_{\varphi}=\frac{\partial {\cal L}}{\partial \dot{\varphi}}
=\dot{\varphi}, \;\;\;
\pi_{\chi}=\frac{\partial {\cal L}}{\partial \dot{\chi}}
=\frac{e}{v} (v+\varphi)^2 \bar{A}_0 \]
and their equal-time commutation relations are
\bea
\ba{lll}
\label{comm3}
& [\varphi({\bf x}), \pi_{\varphi}({\bf y})]=
[\chi({\bf x}), \pi_{\chi}({\bf y})]=
i\delta^2({\bf x}-{\bf y}), \\
& [\varphi({\bf x}), \pi_{\chi}({\bf y})]=
[\chi({\bf x}), \pi_{\varphi}({\bf y})]=0, \\
& [\varphi({\bf x}), \chi({\bf y})]=
[\pi_{\varphi}({\bf x}), \pi_{\chi}({\bf y})]=0.
\ea 
\eea   
The commutation relations of gauge fields are identical with those in 
the symmetric phase. 

Note that the gauge symmetry still remains in the 
symmetry-broken phase as long as we keep the would-be-Goldstone 
boson $\chi$ in the Lagrangian (\ref{lag2}) except the gauge fixing term. 
The supplementary conditons in the symmetry-broken 
phase correspond to this gauge symmetry. 
They are also equal to Eqs. (\ref{F1}) 
and (\ref{G1}), except that $J_0({\bf x})$ is given by 
\be
\label{rho1}
J_0({\bf x})=-m\pi_{\chi}({\bf x}).
\ee
According to these supplementary conditions, one can find that 
the variables ${\cal A}^i$, ${\cal E}^i$, $\varphi$, $\pi_{\varphi}$, 
$A^L_i \equiv \partial_i(V+\chi/m)$, 
and $\pi_{A^L_i} \equiv m\partial_i\pi_{\chi}/
(-\nabla^2+m^2)+ m^2\partial_i(U-A_0)/(-\nabla^2+m^2)$ are the only 
independent physical variables, i.e., commute with $F$ and $G$. 

Consider the symmetric 
energy-momentum tensor defined by Eq. (\ref{emt})
\bea 
T_{\mu\nu} =&-&F_{\mu\lambda}{F_{\nu}}^{\lambda} 
+\partial_{\mu}\varphi\partial_{\nu}\varphi
+ e^2 (v+\varphi)^2 \bar{A}_{\mu}\bar{A}_{\nu} \nonumber \\
&-&g_{\mu\nu} \left\{-\frac{1}{4}F_{\lambda\rho}F^{\lambda\rho}
 +\frac{1}{2}(\partial_{\lambda}\varphi)^2+\frac{1}{2}e^2 
(v+\varphi)^2\bar{A}_{\lambda}^2-V(v,\varphi) \right\}.\label{emt1}
\eea 
As in the symmetric phase, we have dropped the gauge fixing 
term ${\cal L_{GF}}$ from the 
Lagrangian (\ref{lag2}) since we will deal with only 
physical variables commuting with the gauge fixing term. 
According to Eq. (\ref{emt1}), for a small excitation $\varphi$, 
the Hamiltonian is 
\bea
H \approx && \frac{1}{2} \int d^2 x \left\{{\cal E}_i^2 
+{\cal B}^2+\pi^{2}_{\varphi}+(\nabla \varphi)^2 +
\pi^{2}_{\chi}+ m^2(A_i^L)^2
+m^2{\cal A}_i^2 \right\} \nonumber \\
&-& \frac{1}{2}\int d^2 x d^2 y J_0({\bf x})
G({\bf x} - {\bf y})J_0({\bf y})+ H_{int}+ V(v,\varphi),\nonumber \\
=&& \frac{1}{2} \int d^2 x \left\{{\cal E}_i^2 
+{\cal B}^2+\pi^{2}_{\varphi}+(\nabla \varphi)^2 
+ \pi_{A^L_i}(1-\frac{\nabla^2}{m^2})\pi_{A^L_i}
+ m^2(A_i^L)^2 +m^2 {\cal A}_i^2\right\} 
+ H_{int}+ V(v,\varphi),
\eea
where ${\cal B} \equiv \epsilon^{ij} \partial_i {\cal A}^j $. 
The Coulomb-like energy appears as the result of applying the 
supplementary condition (\ref{G1}) with the Eq. (\ref{rho1}). 
After a canonical transformation, which is a Bogoliubov transformation 
with respect to $A_i^L$ defined by $\tilde\pi_{A^L_i}=
\sqrt{1-\frac{\nabla^2}{m^2}}\pi_{A^L_i}$ and 
$\tilde A^L_i= A^L_i/ \sqrt{1-\frac{\nabla^2}{m^2}}$, 
$H$ finally becomes
\be
H=\frac{1}{2} \int d^2 x \left\{{\cal E}_i^2 
+{\cal B}^2+\pi^{2}_{\varphi}+(\nabla \varphi)^2 
+ {\tilde\pi_{A^L_i}}^2 +(\nabla\tilde A_i^L)^2
+m^2 {\cal A}_i^2+m^2 (\tilde A_i^L)^2 \right\}
+ H_{int}+ V(v,\varphi).
\ee
This result confirms that the vector fields are excitations 
with mass $m$.

However, note that the following expression about the 
angular momentum $L$ does not show the result of Binegar 
\cite{Bine} that a massive vector field is spin 1:
\bea
\label{sbL}
L & =& \int d^2 y \epsilon^{ij} y_i T_{0j} \nonumber \\
& = & \int d^2 y \epsilon^{ij} y_i\left\{\pi_{\varphi}({\bf y})
\partial_{j}\varphi({\bf y}) + \dot{\cal A}_k({\bf y}) \partial_j
{\cal A}_{k}({\bf y})
+\pi_{A^L_k}({\bf y})\partial_j A^L_k ({\bf y})\right\},
\eea
where we have dropped would-be-spin terms on vector fields 
since they are total derivatives and vanishe at spatial infinity. 
Notice that, for massive theory, the transverse-longitudinal 
decompositions, Eqs. (\ref{tlda}) and (\ref{tlde}), are ambiguous 
in zero-momemtum limit \cite{Dese82}. Since the spin of 
one-particle states can be characterized by the value of the angualar 
momentum of the particle about an arbitrary axis when the particle 
is at rest, it is not good prescription for the spin 
contents of massive vector fields to deal with the transverse 
and longitudinal components separately. 
Of course, as considered in Ref. \cite{Dese82}, after the removal 
of an infrared singularity of boost generators, 
one can fix the spin of vector excitations. 
In Ref. \cite{LeeM}, the determination of the spin of massive 
vector fields was already performed on along this line, where 
it was shown that the massive vectors carry spin 1. 
This result implies that there exists a spin phase transition in the
three dimensional Maxwell-Higgs theory. 

If now the would-be-Goldstone 
boson $\chi$ is gauged away, there is no supplementary condition 
because there is no gauge symmetry.
The canonical variables $A_0$ and $\pi_0$ are removed through 
Dirac bracket since the constraints, $\pi_0 \approx 0$ and 
$\dot{\pi}_0 \approx 0$, are second-class \cite{Dirac64,Gitm90}.
Thus, there is no need to explicitly specify physical 
variables as the field algebra includes only observable fields, 
$A_1$ and $A_2$, as already confirmed. 
In addition, if we would not take the transverse-longitudinal 
decomposition of the massive vector fields which is observer-dependent, 
the two degrees of freedom will be available to form parity doubled, 
non-zero spin states. 
Then, it can be shown with no wonder the Poincar\'e algebra is well-defined 
and so the spin contents of vector fields can be extracted from it 
with no ambiguity. We will perform this in Appendix.
  
\section{Anyonic physical observables and spin phase transition}   

\subsection{Symmetric Phase}

In $(2+1)$-dimensional Maxwell theory, we showed 
that the fields have no rotational anomaly in any phase, so that 
there is no exotic statistics. But this story is dramatically changed 
in Chern-Simons theory. 
The model we wish to analyze is Chern-Simons gauge theory coupled to 
complex scalar fields in the covariant gauge \cite{mrfj,shin-brst}. 
The Lagrangian density in a symmetric phase is given by
\bea
\label{lag3}
\ba{ll}    
& {\cal L} = \frac{\kappa}{4}\epsilon^{\mu\nu\lambda}A_{\mu}F_{\nu\lambda}
+ |D_{\mu}\phi|^2+{\cal L_{GF}},\\ 
& {\cal L_{GF}} = -\frac{1}{2}(\partial_{\mu}A^{\mu})^2.
\ea
\eea
In spite of gauge fixing term, there are primary constraints given by 
\be
\label{const}
\pi^i-\frac{\kappa}{2} \epsilon^{ij} A_j \approx 0, \;\;\;i,j=1,2,
\ee 
which are second-class constraints which no longer lead to secondary 
constraints. We shall follow the Dirac procedure to 
eliminate these constraints. 
We quantize the theory canonically by introducing 
Dirac brackets in the standard manner 
\cite{Dirac64,Gitm90}.   
The nonvanishing set of equal-time commutation relations is
\bea
\label{comm4}
\ba{lll}
& [A_0({\bf x}), \pi_0({\bf y})] =
i\delta^2({\bf x} - {\bf y}), \\
& [A_i({\bf x}), A_j({\bf y})] =
\frac{i}{\kappa}\epsilon_{ij}\delta^2({\bf x} - 
{\bf y}), \\
& [\phi({\bf x}),\pi({\bf y})] =
[\phi^{\ast}({\bf x}),\pi^{\ast}({\bf y})] =
i\delta^{2}({\bf x} - {\bf y}).
\ea
\eea

As in the Maxwell theory, let us decompose the gauge field 
$A_i(\bf x)$ into transverse and longitudinal parts,
\[ A_i({\bf x})={\cal A}_i({\bf x})+\frac{\partial V}{\partial x^i} 
=\epsilon_{ij}\frac{\partial U}{\partial x^j}
+\frac{\partial V}{\partial x^i}.\]
Then one can show that the fields $U(\bf x)$ and $V(\bf y)$ satisfy the 
following commutation relation using the second relation 
in Eq. (\ref{comm4}):
\be
\label{uv3}
[U({\bf x}),V({\bf y})]=-\frac{i}{\kappa} G(\bf x - \bf y). 
\ee
As disscussed in Maxwell theory, we must impose the supplementary conditions 
on any state $|P>$ in order to return to the consistent original theory 
from the modified Lagrangian (\ref{lag3}). 
All the supplementary conditions affecting 
dynamical variables at one instant of time are
\bea
\label{F2} 
&&F\equiv\partial_{\mu}A^{\mu}=\partial_0 A_0-\nabla^2 V\approx 0,\\
\label{G2}
&&G\equiv-\kappa\epsilon^{ij}\partial_i A_{j}+J_0 =\kappa\nabla^2 U
+J_0 \approx 0,
\eea
where charge density $J_0(\bf x)$ is given by
\[ J_0({\bf x})=ie\left\{\pi^{\ast}({\bf x})\phi^{\ast}({\bf x})- 
\pi({\bf x})\phi({\bf x})\right\}. \] 
Of course, $F$ and $G$ commute with each other. 
Then any components of Chern-Simons gauge field $A_{\mu}$ or any combinations 
of them are not physical, i.e., do not commute with $F$ and $G$. 
Even any component of the field tensor 
$F_{\mu\nu}$ is not physical, although it is gauge invariant. 
As is well-known \cite{Hage84}, there exists no real photon 
in pure Chern-Simons theory of symmetric phase.

The variable $\phi$ is not physical since it commutes with 
$F$ but not with $G$. However, if we define \cite{Dirac58}   
\be
\label{dphi3}
\tilde\phi(x)=e^{ieV(x)}\phi(x),
\ee
$\tilde\phi(x)$ commutes with $F$ and $G$ and thus physical. Likely, 
$\tilde\phi^{\ast}(x)$ is physical. 
Then, the variables $\tilde\phi(x)$, $\tilde \pi(x)$, 
$\tilde\phi^{\ast}(x)$, and $\tilde\pi^{\ast}(x)$ are the only 
independent physical variables, apart from the 
quantities (\ref{F2}) and (\ref{G2}). 
From the commutation relations (\ref{comm4}) and (\ref{uv3}), 
we observe that these physical field operators $\tilde\phi(\bf x)$ 
and $\tilde\phi^{\ast}(\bf x)$ create 
a flux quantum as well as a $U(1)$ charge,
\bea
\ba{ll}
\label{qbr}
Q\left\{\tilde\phi({\bf x})|P>\right\}
=(q-e)\tilde\phi({\bf x})|P>,\;\;\; & \Phi\left\{\tilde\phi({\bf x})|P>\right\}
=(b+\frac{e}{\kappa})\tilde\phi({\bf x})|P>,\\
Q\left\{\tilde\phi^{\ast}({\bf x})|P>\right\}
=(q+e)\tilde\phi^{\ast}({\bf x})|P>,\;\;\; 
& \Phi\left\{\tilde\phi^{\ast}({\bf x})|P>\right\}=
(b-\frac{e}{\kappa})\tilde\phi^{\ast}({\bf x})|P>, 
\ea
\eea
where $Q=\int d^2 x J_0({\bf x}),\;\Phi=\int d^2 x B({\bf x})$ and
a state $|P>$ is assumed to be a simultaneous eigenstate of 
$Q$ and $\Phi$ with eigenvalues $q$ and $b$, respectively.  
These properties show a manifest evidence that 
Chern-Simons gauge fields attach a flux quantum proportional to 
$U(1)$ charge to complex scalar field, which is a dual picture that 
a physical electron in Maxwell theory carries 
Coulomb fields surrounding the charge. 
In the presence of Chern-Simons gauge fields, 
the complex scalar field, $\tilde\phi(\bf x)$ 
or $\tilde\phi^{\ast}(\bf x)$, becomes a boson plus flux 
composite and, as we will see, this composite dynamically will be 
an anyon by the Aharonov-Bohm effect \cite{Ahar59}. 
That is, if we interchange the two field quanta with charge $q$, 
the statistical phase by the 
Aharonov-Bohm effect is equal to $q^2/2\kappa$. 
Indeed, we will construct the composite anyon operators satisfying 
fractional statistics \cite{Seme88} under a covariant gauge fixing.

One can solve the Gauss-law constraint in Fock space represented by the 
physical variables which satisfy supplementary 
conditions (\ref{F2}) and (\ref{G2}): 
\bea
\label{U}
&& U({\bf x})=-\frac{1}{\kappa}\int d^2 y G({\bf x} - {\bf y})J_0({\bf y}),\\ 
\label{V}
&& V({\bf x})=\int d^2 y G({\bf x} - {\bf y})\partial_0 A_{0}({\bf y}).
\eea
The gauge field $A_i (\bf x)$ can be thus expressed as the following 
combination:
\[ A_i({\bf x})=-\frac{1}{\kappa}\epsilon_{ij}\partial_j 
\int d^2 y G({\bf x} - {\bf y})J_0({\bf y}) + 
\partial_i\int d^2 y G({\bf x} - {\bf y})\partial_0 A_{0}({\bf y}).\]
 
The angular momentum operator obtained from symmetric energy-momentum 
tensor given by 
\[T_{\mu\nu} =(D_{\mu}\phi)^*D_{\nu}\phi+(D_{\nu}\phi)^*D_{\mu}\phi
-g_{\mu\nu} |D_{\lambda}\phi|^2\] 
is represented as follows
\bea
\label{ang1}
L &  =& \int d^2 y \epsilon^{ij} y_i T_{0j} \nonumber \\
& = & \int d^2 y \epsilon^{ij} y_i\left\{ \tilde\pi({\bf y})
\partial_{j}\tilde\phi({\bf y}) + \tilde\pi^{\ast}({\bf y})
\partial_{j}\tilde\phi^{\ast}({\bf y})\right\}-\int d^2 y \epsilon^{ij}
y_i {\cal A}_{j}({\bf y})J_0({\bf y}) \nonumber \\ 
& = & \int d^2 y \epsilon^{ij} y_i \left\{\tilde\pi({\bf y})
\partial_{j}\tilde\phi({\bf y}) + \tilde\pi^{\ast}({\bf y})
\partial_{j}\tilde\phi^{\ast}({\bf y})\right\} 
+\frac{Q^2}{4\pi\kappa},
\eea
where we used the equation (\ref{U}) in the final step. 
For the same reason as in Sec.II, we have safely dropped 
the gauge fixing term.
First, note that physical and gauge invariant scalar fields  
$\tilde\phi(x)$ and $\tilde\phi^{\ast}(x)$ have an anomalous spin. 
That is, 
\bea
\label{cang4}
[L, \tilde\phi({\bf x})]&=& 
-i\epsilon^{ij} x_i \partial_{j}\tilde\phi({\bf x}) - 
\int d^2 y \epsilon^{ij} y_i [{\cal A}_{j}({\bf y})J_0({\bf y}),
\tilde\phi({\bf x})] \nonumber \\
&=& -i\epsilon^{ij} x_i \partial_{j}\tilde\phi({\bf x}) + 
\frac{e}{\kappa}\int d^2 y (x_i \partial^{x}_i+y_i \partial^{y}_i) 
G({\bf x} - {\bf y})J_0({\bf y})\tilde\phi({\bf x}) \nonumber \\
&=& -i\epsilon^{ij} x_i \partial_{j}\tilde\phi({\bf x}) - 
\frac{eQ}{2\pi\kappa} \tilde\phi({\bf x}),          
\eea
where we have computed the commutator using the 
relations (\ref{comm4}) and (\ref{uv3}) and used 
the expression (\ref{U}) after this computation.
Similarly,
\be
\label{cang5}
[L, \tilde\phi^{\ast}({\bf x})]= 
-i\epsilon^{ij} x_i \partial_{j}\tilde\phi^{\ast}({\bf x}) + 
\frac{eQ}{2\pi\kappa} \tilde\phi^{\ast}({\bf x}).
\ee
These results agree with the previous ones \cite{Hage84} 
under Coulomb gauge.  

In symmetric phase, the Chern-Simons gauge fields are by themselves 
non-dynamical as the result of ``too much symmetry'' - diffeomorphism 
invariance. As the result of this ``too much symmetry'', 
these gauge fields remain confined, but change a 
boundary condition of coupled fields in the same way that 
the unphysical fields of Maxwell theory, i.e., 
scalar and longitudinal photons, result in an infrared dressing of 
static Coulomb field to charged matter fields \cite{Dirac58}. 
So there can be two points of 
view describing the charged matter system coupled 
to Chern-Simons gauge fields. One is that a part of dynamical 
informations (boundary condition of the coupled fields) assigns to 
the Chern-Simons gauge fields through an interaction. The other is 
that all the dynamical informations are assigned to the charged matter 
fields by removing the Chern-Simons gauge fields through a singular 
gauge transformation initiated by Semenoff \cite{Seme88}. 
However, in quantum field theory described by {\it smooth} fields, 
the removing of a topological term with diffeomorphism invariance 
- Chern-Simons term - through 
singular gauge transformations is in general impossible 
and instead remains a remnant \cite{Jack90}. 
This phenomenon seems to be a quite general feature of bosonization 
in a continuum field theory in higher dimension $D \geq 3$. 

The presence of non-dynamical Chern-Simons gauge fields 
leads a {\it scalar field} to anomalous spin term.
In order to incorporate the relation between spin and statistics, 
we shall construct anyon operators satisfying graded 
commutation relations showing the fractional statistics \cite{Seme88}. 
This is based on the fact that the choice {\it \`a la} Semenoff 
with respect to the physical variables about complex fields $\phi$ 
and $\phi^{\ast}$ is also true:
\bea
\label{hat1}
\hat{\phi}(x) &=& e^{\left\{i\frac{e}{2\pi\kappa}\int d^2 y 
\Theta({\bf x} - {\bf y})J_0({\bf y})
+ieV({\bf x})\right\}}\phi(x) \nonumber \\ 
&\equiv& S(x)\phi(x), \\
\label{hat2}
\hat{\phi}^{\ast}(x) &=& e^{\left\{-i\frac{e}{2\pi\kappa}\int d^2 y 
\Theta({\bf x} - {\bf y})J_0({\bf y})
-ieV({\bf x})\right\}}\phi^{\ast}(x) \nonumber \\
&\equiv& S^{-1}(x)\phi^{\ast}(x).
\eea
We introduced the multivalued function $\Theta({\bf x})$ satisfying 
``Cauchy-Riemann'' equations
\be
\label{CR}
\partial_i G({\bf x})=\frac{1}{2\pi}\epsilon_{ij}\partial_j 
\Theta({\bf x})
\ee
and the resulting function $\Theta({\bf x})$ satisfies 
the following properties
\bea
\label{theta}
\ba{ll}
& \mbox{tan}\Theta({\bf x})=\frac{x_2}{x_1},\;\;\; 
\partial_i \Theta({\bf x})=
-\epsilon_{ij}\frac{x^j}{{\bf x}^2},\\
& \nabla \times \nabla \Theta({\bf x})=2\pi\delta^2({\bf x}),\;\;\; 
\nabla^2\Theta({\bf x})=0.
\ea 
\eea

Now we want to see the properties of the {\it hat} fields through careful 
analysis. First, let us express $\pi(x)$ in terms of the hat fields.
\begin{eqnarray*} 
 \pi(x) &=& (D_0 \phi(x))^{\ast}=\left\{ (\partial_0+ieA_0)S^{-1}(x)
\hat{\phi}(x)\right\}^{\ast}\\
& =& S(x)\biggr[\left\{\partial_0-ie\int d^2 y 
G({\bf x} - {\bf y})\Box A_0({\bf y})
-i\frac{e}{2\pi\kappa}\int d^2 y \Theta({\bf x} - {\bf y}) 
\partial_{i} J_{i} ({\bf y})\right\}
\hat{\phi}({\bf x})\biggr]^{\ast}, 
\end{eqnarray*}
where we used the Eq. (\ref{V}) and current conservation law. After using 
the equation of motion about $A_0$ and integration by part, we arrive at 
the following result
\bea
\label{hat3}
& &S(x)\biggr[\left\{\partial_0-i\frac{e}{\kappa}\int d^2 y 
\epsilon_{ij}\partial_{i}^{y}G({\bf x} - {\bf y})J_{j}({\bf y})
-i\frac{e}{2\pi\kappa}\int d^2 y \Theta({\bf x} - {\bf y})
\partial_i^{y}J_{i}({\bf y})\right\}
\hat{\phi}({\bf x})\biggr]^{\ast},\nonumber \\
& & =S(x)\biggr[\left\{\partial_0
-i\frac{e}{2\pi\kappa}\int d^2 y \partial_i^{y} \left(\Theta({\bf x} 
- {\bf y})J_{i}({\bf y})\right)\right\}
\hat{\phi}({\bf x})\biggr]^{\ast},\nonumber \\
& & =S(x)\left\{\partial_0 \hat{\phi}(x)\right\}^{\ast}
\equiv S(x)\hat{\pi}(x),
\eea  
where we used ``Cauchy-Riemann'' equation (\ref{CR}) 
and assumed that current density rapidly decreases at large $r$.
In the same way, we can show that
\be
\label{hat4}
\pi^{\ast}(x)=S^{-1}(x)\left\{\partial_0 \hat{\phi}(x)\right\}
\equiv S^{-1}(x)\hat{\pi}^{\ast}(x). 
\ee
Second, express $D_i\phi(x)$ in terms of the hat fields: 
\bea 
\label{chat1}
D_i\phi(x) &=&(\partial_i+ieA_i)\left\{S^{-1}(x)
\hat{\phi}(x)\right\} \nonumber \\
&=&S^{-1}(x)\partial_i\hat{\phi}(x)+ieS^{-1}(x) \left\{{\cal A}_i(x)
-\frac{1}{2\pi\kappa}\partial_i\int d^2 y \Theta({\bf x} - {\bf y})
J_0({\bf y})\right\}\hat{\phi}(x).
\eea
As a consequence of Eqs. (\ref{U}) and (\ref{CR}), 
${\cal A}_i(x)$ can be rewritten as
\be
\label{cala1}
{\cal A}_i(x)=\frac{1}{2\pi\kappa}\int d^2 y\partial_i^{x} 
\Theta({\bf x} - {\bf y})J_0({\bf y}).  
\ee
However, unless the charge density $J_0({\bf y})$ is sufficiently well 
localized, the interchange of integral and derivative in Eq. (\ref{cala1}) 
and thus displaying ${\cal A}_i(x)$ as a pure gauge is in general not 
correct \cite{Jack90}. When $J_0({\bf y})$ is smoothly distributed over an 
extended region, the correct expression for ${\cal A}_i$ is 
\be
\label{cala2}
{\cal A}_i({\bf x}) = \frac{1}{2\pi\kappa}\partial_i\int d^2 y 
\Theta({\bf x} - {\bf y})J_0({\bf y})-\frac{1}{2\pi\kappa}
\epsilon_{ij}\int d\theta d{\bf z}^j_{\theta}J_0({\bf x};{\bf z}),  
\ee     
where the line integral of $J_0$ is along the cut line introduced to 
integrate the multivalued function $\Theta$ and $\theta$ is the polar 
angle of the cut line. In order to remove an arbitrary directional 
dependence of minimal coupling arising from the choice of cut-line, 
we considered the averaging process for all possible choices 
of cut-line. This expression is still consistent with 
Gauss-law constraint $B=\nabla \times {\bf A}=-\frac{1}{\kappa}J_0$. 
For the sufficiently well localized charge density such as nonrelativistic 
quantum field theory \cite{Jack90} or theory on the lattice \cite{Frad89}, 
the second term in Eq. (\ref{cala2}) can be dropped with impunity. 
Anyway, if we can {\it neglect} the 
contribution of the remnant in Eq. (\ref{cala2}) for some 
distribution $J_0$, we finally arrive at the following result:
\[D_i\phi(x) =S^{-1}(x)\partial_i\hat{\phi}(x),\;\;\;
(D_i\phi(x))^{\ast} =S(x)\partial_i\hat{\phi}^{\ast}(x).\]  
The Hamiltonian of hat fields then takes the form
\be
\l{H}
H =\int d^2 x \left\{ \frac{1}{2} \hat{\pi}^{2}+
\frac{1}{2}(\nabla \hat{\phi})^2 \right\},
\ee
and their angular momentum operator is given by
\be
\l{L}
L=\int d^2 x \epsilon^{ij} x_i \left\{\hat{\pi}\partial_{j}\hat{\phi}+ 
\hat{\pi}^{\ast}\partial_{j}\hat{\phi}^{\ast}\right\}.
\ee

To study the statistics of hat fields, we use the identities by the 
Baker-Campbell-Hausdorff formula,
\bea
\label{HB}
&&S(x)\phi(z)S^{-1}(x)=e^{-i\frac{e^2}{2\pi\kappa}
\Theta({\bf x} - {\bf z})}\phi(z), \nonumber \\ 
&&S(x)\pi(z)S^{-1}(x)=e^{i\frac{e^2}{2\pi\kappa}
\Theta({\bf x} - {\bf z})}\pi(z). 
\eea
The commutation relations of hat fields now obey the graded commutation 
relations \cite{Seme88},
\bea
\label{gcr}
& & \hat{\phi}({\bf x})\hat{\phi}({\bf y})=e^{-i\frac{e^2 \Delta}{2\pi\kappa}}
\hat{\phi}({\bf y})\hat{\phi}({\bf x}), \nonumber \\ 
& & \hat{\phi}({\bf x})\hat{\phi}^{\ast}({\bf y})=
e^{i\frac{e^2 \Delta}{2\pi\kappa}}
\hat{\phi}^{\ast}({\bf y})\hat{\phi}({\bf x}), \nonumber \\
& & \hat{\phi}({\bf x})\hat{\pi}({\bf y})=i\delta^2({\bf x}-{\bf y})+
e^{i\frac{e^2 \Delta}{2\pi\kappa}}\hat{\pi}({\bf y})
\hat{\phi}({\bf x}), \\
& & \hat{\phi}({\bf x})\hat{\pi}^{\ast}({\bf y})=
e^{-i\frac{e^2 \Delta}{2\pi\kappa}}\hat{\pi}^{\ast}({\bf y})
\hat{\phi}({\bf x}), \nonumber \\
& & \hat{\pi}({\bf x})\hat{\pi}({\bf y})=e^{-i\frac{e^2 \Delta}{2\pi\kappa}}
\hat{\pi}({\bf y})\hat{\pi}({\bf x}), \nonumber \\ 
& & \hat{\pi}({\bf x})\hat{\pi}^{\ast}({\bf y})=
e^{i\frac{e^2 \Delta}{2\pi\kappa}}\hat{\pi}^{\ast}({\bf y})
\hat{\pi}({\bf x}), \nonumber 
\eea
with multi-valued phase
\[\Delta=\Theta({\bf x}-{\bf y})-\Theta({\bf y}-{\bf x})=
\pi\;\mbox{mod} \; 2\pi n.\]
These multi-valued operators carry fractional statistics and 
may be regarded as anyon operators 
since they create a state with arbitray spin 
when acting on a physical state.
The statistical phases in the graded commutation relations (\ref{gcr}) 
are exactly equal to the Aharonov-Bohm phase for the field quanta 
satisfying the Eq. (\ref{qbr}). 
Consequently, we see that the Aharonov-Bohm effect is 
the origin of anyon statistics.  

Note that the representation of 
a physical variable depends on the 
gauge fixing and one need to first find all physical variables 
before turning to dynamics in order to obtain correct results. 
This is the lesson we learn from Dirac \cite{Dirac58}.

\subsection{Symmetry-Broken Phase}

We shall now consider when spontaneous symmetry breaking occurs 
\cite{Wen89}. In the same way as the Maxwell theory, 
we introduce symmetry breaking potential $V(\phi)$ 
and consider the same parameterization of 
a complex scalar field $\phi$. 
Then the Lagrangian density in symmetry-broken phase is given by
\be
\label{lag4}
{\cal L} = \frac{\kappa}{4}\epsilon^{\mu\nu\lambda}A_{\mu}F_{\nu\lambda} +  
\frac{1}{2}(\partial_{\mu}\varphi)^2+\frac{1}{2}e^2 
(v+\varphi)^2(\bar{A}_{\mu})^2 -V(v,\varphi)+{\cal L_{GF}}, 
\ee
where $\bar{A}_{\mu}$ is defined by
\be
\label{bara2}
\bar{A}_{\mu} \equiv A_{\mu}+\frac{1}{ev}\partial_{\mu}\chi.
\ee 
The effective vector action obtained from (\ref{lag4}) is 
\be
\label{sdca}
{\cal L} = \frac{\kappa}{4}\epsilon^{\mu\nu\lambda}A_{\mu}F_{\nu\lambda}
+\frac{1}{2}e^2v^2A_{\mu}^2,
\ee
after a gauge transformation $A_{\mu}\rightarrow A_{\mu}-
\frac{1}{ev}\partial_{\mu}\chi.$
The ``self-dual'' first order system (\ref{sdca}) has been shown 
\cite{Dese84} to be equivalent to topologically massive 
spin 1 theory \cite{Dese82} even in the presence of interaction. 
Thus one expects that the Chern-Simons gauge field absorbed 
a would-be-Goldstone boson is transmuted into topologically massive 
helicity 1 excitation in the Higgs' phase. This is the observation 
of Deser and Yang \cite{Dese89}. 
However, there also exist arguments that 
the excitation relative to ground state 
is spin 0 boson \cite{Boya90} and that the spin phase transition 
is impossible in the Chern-Simons-Higgs theory \cite{shin-ssb}. 
We shall resolve this inconsistency 
based on the same analysis as in Sec.II.
 
The equal-time commutation relations of Chern-Simons gauge fields are 
equal to Eq. (\ref{comm4}) and matter parts are equal to Eq. (\ref{comm3}). 
The supplementary conditions are also identical to Eqs. (\ref{F2}) 
and (\ref{G2}), except that $J_0({\bf x})$ is given by 
\be
\label{rho2}
J_0({\bf x})=-ev\pi_{\chi}({\bf x}).
\ee
According to these supplementary conditions, we can find that 
the variables $\varphi(x)$, $\pi_{\varphi}$, $A^L_i \equiv 
\partial_i (V+ \chi/ev)$, and 
$\pi_{A^L_i} \equiv ev\partial_i\pi_{\chi}/
(-\nabla^2+m^2)+ \kappa m^2\partial_i U/(-\nabla^2+m^2) 
\;(m=e^2v^2/\kappa)$ are the only independent physical variables, 
apart from the 
quantities (\ref{F2}) and (\ref{G2}). Note that the longitudinal 
Chern-Simons gauge field absorbed a would-be-Goldstone boson can be 
a dynamical variable and restore the vector field's 
dynamics \cite{Dese89} in the Higgs' phase. 
That is, the Chern-Simons-Higgs theory will have one massive spin 1 
and one massive spin 0 (Higgs field) propagating modes. 
In the case of Maxwell-Chern-Simons-Higgs theory, 
we again have two parity-violating massive spin 1 modes \cite{LeeM,Paul86}.  
Although the scalar form $V+\chi/ev$ identified 
in Refs. \cite{Boya90} and \cite{shin-ssb} 
is {\it physical} in the Dirac's sense, the more appropriate choice 
will be vector form as seen from the 
experience obtained by the analysis of the Maxwell-Higgs theory in Sec.II. 
In addition, their choice will encounter an infrared 
singularity since $V+\chi/ev=\partial_i A^L_i/
\nabla^2$ and this will bring about a superfluous infrared 
singularity in Poincar\'e algebra.  

In order to examine the physical spectrum of the involved fields, 
consider the symmetric energy-momentum tensor 
defined by Eq. (\ref{emt}):
\be
\label{emt2}
T_{\mu\nu} = \partial_{\mu}\varphi\partial_{\nu}\varphi
+ e^2 (v+\varphi)^2 \bar{A}_{\mu}\bar{A}_{\nu}-
g_{\mu\nu} \left\{ \frac{1}{2}(\partial_{\lambda}\varphi)^2
+\frac{1}{2}e^2(v+\varphi)^2\bar{A}_{\lambda}^2-V(v,\varphi) \right\}.
\ee
Then the Hamiltonian becomes 
\bea
\l{ham}
H &=& \int d^2 x \left\{ \frac{1}{2}\left(\dot{\varphi}^2
+(\nabla \varphi)^2 +
e^2 (v+\varphi)^2 \bar{A}_0\bar{A}_0+e^2 (v+\varphi)^2 \bar{A}_i 
\bar{A}_i\right)+V(v,\varphi) \right\} \nonumber\\
&=& \frac{1}{2} \int d^2 x \left\{\pi^{2}_{\varphi}+(\nabla \varphi)^2 
+ \pi_{A^L_i}(\frac{e^2v^2}{\kappa^2}-\frac{\nabla^2}{e^2v^2})\pi_{A^L_i}
+ e^2v^2(A_i^L)^2 \right\}+ H_{int}+ V(v,\varphi),
\eea
where Coulomb-like energy appears as the result of applying 
the supplementary condition (\ref{G2}). 
As in the Sec.II, after a Bogoliubov transformation with respect to 
$A_i^L$ defined by $\tilde\pi_{A^L_i}=
\sqrt{\frac{e^2v^2}{\kappa^2}-\frac{\nabla^2}{e^2v^2}}\pi_{A^L_i}$ and 
$\tilde A^L_i= A^L_i/ \sqrt{\frac{e^2v^2}{\kappa^2}-
\frac{\nabla^2}{e^2v^2}}$, $H$ becomes
\be
H=\frac{1}{2} \int d^2 x \left\{\pi^{2}_{\varphi}+(\nabla \varphi)^2 
+ {\tilde\pi_{A^L_i}}^2 +(\nabla\tilde A_i^L)^2
+\frac{e^4v^4}{\kappa^2} (\tilde A_i^L)^2 \right\}
+ H_{int}+ V(v,\varphi).
\ee
This result confirms that the vector field is an excitation 
with mass $m=e^2v^2/\kappa$.

The angular momentum operator obtained from the symmetric energy-momentum 
tensor (\ref{emt2}) is represented by using the supplementary condition 
(\ref{G2}) as follows,
\bea
\label{ang2}
L &  =& \int d^2 y \epsilon^{ij} y_i T_{0j} \nonumber \\
& = & \int d^2 y \epsilon^{ij} y_i\left\{\pi_{\varphi}({\bf y})
\partial_{j}\varphi({\bf y}) + ev\pi_{\chi}({\bf y})
A_j^L({\bf x})\right\}-\int d^2 y \epsilon^{ij}
y_i {\cal A}_{j}({\bf y})J_0({\bf y}) \nonumber \\ 
& = & \int d^2 y \epsilon^{ij} y_i\left\{\pi_{\varphi}({\bf y})
\partial_{j}\varphi({\bf y}) + 
\pi_{A^L_k}({\bf y})\partial_j A^L_k ({\bf y})\right\}
+\frac{Q^2}{4\pi\kappa},
\eea
where we have again dropped a would-be-spin term on vector field 
since it is total derivative and vanishes at spatial infinity. 
The angular momentum operator $L$ in Eq. (\ref{ang2}) seems to have 
the anomalous term as in Ref. \cite{Boya90} but it is misleading 
since the last term in Eq. (\ref{ang2}) induces no effect 
on fields, i.e.,
\be
\l{0s}
[\int d^2 y \epsilon^{ij}y_i {\cal A}_{j}({\bf y})J_0({\bf y}), 
A_i^L(x)]
=[\int d^2 y \epsilon^{ij}y_i {\cal A}_{j}({\bf y})J_0({\bf y}), 
\varphi(x)]=0.
\ee 
Moreover, it is expected that, 
in the massive vector theory, we will encounter 
the zero-momentum ambiguity on the transverse-longitudinal 
decomposition of vector fields, which brings about the zero-momentum 
anomaly in Poincar\'e algebra as in Sec.II. 
Thus the correct spin content of vector fields 
should be determined by removing zero-momentum singularity 
of boost generator or the prescription free from zero-momentum 
ambiguity which abandons the transverse-longitudinal decomposition 
of vector fields. 
In the Appendix, we will obtain the result that the Chern-Simons guage
field $A_i^L$ is a spin 1 excitation, 
by checking the Poincar\'e algebra in terms of the prescription 
free from zero-momentum ambiguity which 
does not take the transverse-longitudinal decomposition of vector fields.

Deser and Jackiw found \cite{Dese84} that massive Chern-Simons theory is 
equivalent (by a Legendre transformation) to topologically massive gauge 
theory \cite{Dese82} which has a parity-violating spin 1 excitation, 
where the spin of the vector excitation was fixed by the removal of 
zero-momentum singularity of Poincar\'e algebra. 
In Ref. \cite{Dese89}, Deser and Yang observed that Higgs mechanism 
transmutes a non-dynamical Chern-Simons term into topologically massive, 
parity-violating, spin 1 theory. According to the results, 
we can conclude that the Chern-Simons gauge field in Higgs' phase 
is a spin 1 excitation. 
Consequently, the excitations in Higgs' phase have 
no anomalous spin and the Chern-Simons gauge field absorbed 
a would-be-Goldstone boson is transmuted into topologically massive 
helicity 1 excitation. 
Therefore, the Higgs mechanism transmutes an anyon satisfying 
fractional statistics into a spin 0 or a spin 1 boson and so 
there exists interestingly a statistical 
spin-phase transition as observed by Wen and Zee \cite{Wen89}. 

\section{CONCLUSION}

We have presented the quantization of a charged matter system coupled to 
the Chern-Simons gauge field in the covariant gauge fixing. 
Our approach is based on the Dirac's method performed on 
QED \cite{Dirac58} which provides us definite way of 
identifying physical spectrums free from gauge ambiguity arising 
from the gauge fixing and illustrates the importance of 
consideration of physical field variables. 
The important point is that the formulation of dynamics 
in terms of a Lagrangian (or Hamiltonian) and 
the equations of motion make use of a larger field algebra 
which includes non-observable fields and thus one must find all 
physical variables before turning to dynamics in order to obtain 
correct results. Then it can be shown quite generally that the physical 
charged fields are described by non-local fields carrying static field, 
for example, Coulomb fields for Maxwell theory and magnetic flux for 
Chern-Simons theory. In the case of Chern-Simons theory, we have 
shown that the static field, i.e. the magnetic flux, 
attached to charged matter fields 
is the origin of fractional statistics. 

We have also presented the quantization of Chern-Simons matter system 
in a symmetry-broken phase and ensured that Higgs mechanism 
transmutes a non-dynamical Chern-Simons term into 
topologically massive, parity-violating, spin 1 theory. 
Thus Higgs effect transmutes an anyon satisfying 
fractional statistics into a canonical boson, a spin 0 Higgs boson 
or a topologically massive photon which is a Chern-Simons gauge 
field absorbed a would-be-Goldstone boson. 
In order to identify correct spectrums, 
we have used two altenative and complementary 
prescriptions and found 
the consistent result with Deser and Jackiw on the spin of 
massive vector fields and thus removed an inconsistency 
between Boyanovsky and Deser and Yang. 
Consequently, it implies that the Higgs effect induces 
the statistical spin phase transition in the Chern-Simons-Higgs 
theory as well as Maxwell-Higgs theory. 

We think that the same approach performed in this paper 
will be applied to Maxwell-Chern-Simons theory as well. 
For the Maxwell-Chern-Simons theory, there also exist different 
opinions \cite{Foer89,SeSo} on anyon statistics in symmetric phase.
In the symmetry broken Higgs' phase, this theory have two 
parity-violating massive 
spin 1 photons \cite{LeeM,Paul86} and one spin 0 Higgs field. 
Thus the issue on the existence of anyon statistics in this 
model will be involved with that of 
the statistical spin phase transition. 
Study on these issues using the same approach as in this paper 
will be also interesting. 

\section*{ACKNOWLEDGEMENTS}
We thank Prof. Sang-Jin Sin for useful discussions. 
This work was supported by the Korean Science and 
Engeneering Foundation through 
the Center for Theoretical Physics and 
by the Korean Ministry of Education (BSRI-98-2414).

\appendix
\section{Poincar\'e algebra of massive vector fields in 2+1 dimensions}
\subsection{Proca theory}
In this appendix, we will analyze the Poincar\'e algebra of Proca theory 
in 2+1 dimensions and show that the massive vector field in the 
Proca theory is canonically spin 1. 

Consider the following Proca Lagrangian
\be
\label{Apl}
{\cal L}_P= -\frac{1}{4}F_{\mu\nu}F^{\mu\nu}+
\frac{1}{2}m^2 {A}_{\mu}^2.
\ee
After the canonical quantization of the theory by introducing 
Dirac bracket, we obtain the following commutation relations:
\bea
\label{Acr}
&& [A_0({\bf x}), \pi_0({\bf y})] =0, \nonumber \\
&& [A_i({\bf x}), \pi^j({\bf y})] =i\delta_{ij}\delta^2({\bf x} - 
{\bf y}),
\eea
where conjugate momenta $\pi^j \equiv \frac{\partial {\cal L}_P}
{\partial {\dot A}_j}=F^{j0}$. 
Then the energy-momentum tensor $T_{\mu\nu}$ for the Proca 
theory (\ref{Apl}) is given as
\be
\label{Apem}
T_{\mu\nu} =-F_{\mu\lambda}{F_{\nu}}^{\lambda} 
+m^2A_{\mu}A_{\nu}-g_{\mu\nu}{\cal L}_P.
\ee
With the energy-momentum tensor, the Poincar\'e generators can be 
expressed as following forms,
\bea
\label{App}
\ba{llll}
& P_i=\int d^2 x T_{0i}=\int d^2 x \pi^k(x)\partial_i A_k(x),\\
& H=\int d^2 x T_{00}=\int d^2 x \{\frac{1}{2}\pi^i(x)\pi^i(x)
+\frac{1}{4}F^{ij}(x)F_{ij}(x)+\frac{1}{2m^2}(\partial_i\pi^i(x))^2+
\frac{m^2}{2}A^i(x)A^i(x)\}, \\  
&L=\int d^2 x \epsilon^{ij} x_i T_{0j}=
\int d^2 x \epsilon^{ij} x_i\pi^k(x)\partial_{j}A_k(x)
-\int d^2 x \epsilon^{ij}\pi^i(x)A_j(x),\\
&M_{i0}=\int d^2 x x^i \{\frac{1}{2}\pi^j(x)\pi^j(x)
+\frac{1}{4}F^{jk}(x)F_{jk}(x)+\frac{1}{2m^2}(\partial_j\pi^j(x))^2+
\frac{m^2}{2}A^j(x)A^j(x)\}-tP_i,
\ea
\eea
using the equation of motion, $A_0=-\frac{1}{m^2}\partial_i F^{i0}
=-\frac{1}{m^2}\partial_i \pi^i$. 
After a straightforward calculation with the commutation 
relations (\ref{Acr}), one can check that the Poincare algebra for 
the Proca theory is well-defined, especially,
\bea
\label{Ahl}
\ba{ll}
&[M_{i0}, P_j]=-i\delta_{ij} H, \\
&[M_{i0}, M_{j0}]=-i\epsilon_{ij}L.
\ea
\eea

The Poincar\'e algebra is free from zero-momentum anomaly and 
the angular momentum operator $L$ in Eq. (\ref{App}) has 
a canonical expression for spin 1 theory contrary to 
that of Eq. (\ref{sbL}).  
Here, we confirm the result of Binegar \cite{Bine} that 
a massive vector field is spin 1.

\subsection{Massive Chern-Simons theory}

In this appendix, we will analyze the Poincar\'e algebra of 
massive Chern-Simons theory in 2+1 dimensions which has only one 
propagating mode and show that this massive mode in the 
Chern-Simons theory is canonically spin 1. 

Consider the following massive Chern-Simons Lagrangian \cite{Dese84}
\be
\label{Acsl}
{\cal L}_ {CS}=\frac{\kappa}{4}\epsilon^{\mu\nu\lambda}A_{\mu}
F_{\nu\lambda}+\frac{\mu}{2}A_{\mu}^2.
\ee
After the canonical quantization of the theory by introducing 
Dirac bracket, we obtain the following commutation relations:
\bea
\label{Acrc}
&& [A_0({\bf x}), \pi_0({\bf y})] =0, \nonumber \\
&& [A_i({\bf x}), A_j({\bf y})] =\frac{i}{\kappa}
\epsilon_{ij}\delta^2({\bf x} - {\bf y}).
\eea
Eq. (\ref{Acrc}) shows that the Chern-Simons gauge fields 
$A_1$ and $A_2$ are not independent 
due to the symplectic structure of ${\cal L}_{CS}$. 

The energy-momentum tensor $T_{\mu\nu}$ for the massive Chern-Simons 
theory (\ref{Acsl}) is given by
\be
\label{Acsem}
T_{\mu\nu} = \mu A_{\mu}A_{\nu}-\frac{\mu}{2}g_{\mu\nu}
A_{\lambda}A^{\lambda}.
\ee
With the energy-momentum tensor, the Poincar\'e generators can be 
expressed as following forms,
\bea
\label{Apcs}
\ba{llll}
& P_i=\int d^2 x T_{0i}=\int d^2 x \kappa B(x)A_i(x)=
\int d^2 x \pi^k(x)\partial_i A_k(x),\\
& H=\int d^2 x T_{00}=\int d^2 x \{\frac{{\kappa}^2}{2\mu}(B(x))^2
+\frac{\mu}{2}A^i(x)A^i(x)\}, \\  
&L=\int d^2 x \epsilon^{ij} x_i T_{0j}=
\kappa\int d^2 x \epsilon^{ij} x_i B(x)A_j(x)=
\int d^2 x \epsilon^{ij} x_i\pi^k(x)\partial_{j}A_k(x)
-\int d^2 x \epsilon^{ij}\pi^i(x)A_j(x),\\
&M_{i0}=\int d^2 x x^i \{\frac{{\kappa}^2}{2\mu}(B(x))^2
+\frac{\mu}{2}A^j(x)A^j(x)\}-tP_i,
\ea
\eea
where $B(x)=\epsilon^{ij}\partial_i A^j(x),\;
\pi^i(x)=\frac{\kappa}{2} \epsilon^{ij} A_j(x)$ and 
$A_0=\frac{\kappa}{\mu}B(x)$. 
After a straightforward calculation using the commutation 
relations (\ref{Acrc}) and $[B(x), B(y)]=0$, 
one can also check that the Poincare algebra for 
the massive Chern-Simons theory is well-defined, especially,
\bea
\label{Achl}
\ba{ll}
&[M_{i0}, P_j]=-i\delta_{ij} H, \\
&[M_{i0}, M_{j0}]=-i\epsilon_{ij}L.
\ea
\eea

The Poincar\'e algebra is also free from zero-momentum anomaly and 
the angular momentum operator $L$ in Eq. (\ref{Apcs}) has 
a canonical expression for spin 1 theory.  
Thus, we confirm the result of Deser and Jackiw \cite{Dese84} that 
the excitation of the massive Chern-Simons theory is spin 1.


\begin{references}
\bibitem{Hage84}C. R. Hagen, Ann. Phys. (N.Y.) {\bf 157}, 342 (1984); 
Phys. Rev. D {\bf 31}, 2135 (1985). 
\bibitem{WZ83}F. Wilczek and A. Zee, Phys. Rev. Lett. {\bf 51}, 2250 (1983);
Y. -S. Wu and A. Zee, Phys. Lett. B{\bf 147}, 325 (1984); M. J. Bowick, 
D. Karabali, and L. C. R. Wijewardhana, Nucl. Phys. {\bf B271}, 417 (1986).
\bibitem{Poly88}A. M. Polyakov, Mod. Phys. Lett. A {\bf 3}, 325 (1988); 
P. K. Panigrahi, S. Roy, and W. Scherer, Phys. Rev. Lett. 
{\bf 61}, 2827 (1988).
\bibitem{Seme88}G. W. Semenoff, Phys. Rev. Lett. {\bf 61}, 517 (1988).
\bibitem{Forte}S. Forte, Rev. Mod. Phys. {\bf 64}, 
193 (1992); {\it Fractional Statistics and Anyon Superconductivity}, 
edited by F. Wilczek (World Scientific, Singapore, 1990) 
and references therein.
\bibitem{Laug83}R. B. Laughlin, Phys. Rev. Lett. {\bf 50}, 1395 (1983); 
B. I. Halperin, {\it ibid.} {\bf 52}, 1583 (1984); 
{\it The Quantum Hall Effect}, edited by M. Stone (World Scientific, 
Singapore, 1992) and references therein.
\bibitem{Hald85}F. D. M. Haldane and Y.-S. Wu, Phys. Rev. Lett. {\bf 55}, 
2877 (1985); E. Fradkin, {\it Field Theories of Condensed Matter Systems} 
(Addison-Wesley, Redwood City, California, 1991).
\bibitem{Science}R. B. Laughlin, Science {\bf 242}, 525 (1988); 
V. Kalmeyer and R. B. Laughlin, Phys. Rev. Lett. {\bf 59}, 2095 (1987); 
R. B. Laughlin, {\it ibid.} {\bf 60}, 2677 (1988). 
\bibitem{Hage89}C. R. Hagen, Phys. Rev. Lett. {\bf 63}, 1025 (1989); 
G. W. Semenoff, {\it ibid.}, {\bf 63}, 1026 (1989); C. R. Hagen, 
Phys. Rev. D {\bf 44}, 2614 (1991); T. Matsuyama, 
{\it ibid.}, {\bf 44}, 2616 (1991); C. R. Hagen, Phys. Rev. Lett. 
{\bf 70}, 3518 (1993); R. Banerjee, {\it ibid.}, {\bf 70}, 3519 (1993).
\bibitem{Foer89}A. Foerster and H. O. Girotti, Phys. Lett. B{\bf 230}, 
83 (1989); Nucl. Phys. {\bf B342}, 680 (1990).
\bibitem{Bane92}R. Banerjee, Phys. Rev. Lett. {\bf 69}, 17 (1992); 
Phys. Rev. D {\bf 48}, 2905 (1993).
\bibitem{Boya92}D. Boyanovsky, E. T. Newman, and C. Rovelli, 
Phys. Rev. D {\bf 45}, 1210 (1992).
\bibitem{park2}Mu-In Park and Young-Jai Park, hep-th/9803208, 
to appear in Rapid Communications-PRD/Nov/98. 
\bibitem{Dirac58}P. A. M. Dirac, {\it The Principles of Quantum Mechanics} 
(4th ed., London: Oxford Univ. Press, 1958).. 
\bibitem{Strocchi}F. Strocchi, {\it Selected Topics on the General 
Properties of Quantum Field Theory} (World Scientific, Singapore, 1993).
\bibitem{mrfj}M. Mintchev and M. Rossi, Phys. Lett. B{\bf 271}, 187
(1991); S. Forte and T. Jolicoeur, Nucl. Phys. {\bf B350}, 589 (1991). 
\bibitem{shin-brst}H. Shin, W.-T. Kim, J.-K. Kim, and Y.-J. Park, 
Phys. Rev. D{\bf 46}, 2730 (1992); J. K. Kim, W.-T. Kim, and H. Shin, 
J. Phys. A{\bf 27}, 6067 (1994).
\bibitem{Wen89}X. G. Wen and A. Zee, J. Phys. (Paris) {\bf 50}, 1623 (1989).
\bibitem{Boya90}D. Boyanovsky, Phys. Rev. D {\bf 42}, 1179 (1990).
\bibitem{shin-ssb}W.-T. Kim, H. Shin, and J. K. Kim, 
Mod. Phys. Lett. A{\bf 8}, 3317 (1993).
\bibitem{WZ89}X. G. Wen and A. Zee, Phys. Rev. Lett. {\bf 62}, 1937 (1989).
\bibitem{Dese89}S. Deser and Z. Yang, Mod. Phys. Lett. A {\bf 4}, 2123 (1989).
\bibitem{Dese84}S. Deser and R. Jackiw, Phys. Lett. B{\bf 139}, 371 (1984).
\bibitem{Dese82}S. Deser, R. Jackiw, and S. Templeton, Ann. Phys. 
(N.Y.) {\bf 140}, 372 (1982).
\bibitem{Bjor65}J. D. Bjorken and S. D. Drell, {\it Relativistic Quantum 
Fields} (McGraw-Hill, New York, 1965).
\bibitem{Bine}B. Binegar, J. Math. Phys. {\bf 23}, 1511 (1982).
\bibitem{LeeM}B.-H. Lee and H. Min, Phys. Rev. D{\bf 51}, 4458 (1995).
\bibitem{Dirac64}P. A. M. Dirac, {\it Lectures on Quantum Mechanics} 
(Yeshiva Univ. Press, New York, 1964).
\bibitem{Gitm90}D. M. Gitman and I. V. Tyutin, {\it Quantization of Fields 
with Constraints} (Springer-Verlag, Berlin Heidelberg, 1990).
\bibitem{Ahar59}Y. Aharonov and D. Bohm, Phys. Rev. {\bf 115}, 485 (1959).
\bibitem{Jack90}R. Jackiw and S.-Y. Pi, Phys. Rev. D{\bf 42}, 3500 (1990).
\bibitem{Frad89}E. Fradkin, Phys. Rev. Lett. {\bf 63}, 322 (1989).
\bibitem{Paul86}S. K. Paul and A. Khare, Phys. Lett. B{\bf 171}, 244 (1986).
\bibitem{SeSo}G. W. Semenoff and P. Sodano, Nucl. Phys. 
{\bf B328}, 753 (1989); T. Matsuyama, Phys. Rev. D{\bf 42}, 3469 (1990).
\end{references}
\end{document}